\documentclass[11pt]{article} 

\def \noi {\noindent}

\def \ssk{\smallskip} 

\def \aa {{\it Astron. Astrophys.}} 
\def \aj {{\it Astron. J.}}

\def \cmda {{\it Celest. Mech. Dyn. Astr.}}

\usepackage{latexsym}
\usepackage{amsmath}
\usepackage{amssymb}
\usepackage{amsfonts}
\usepackage{graphicx}

\begin{document} 

\vspace*{1.2cm} 
\noi {\Large TEMPORAL VARIATIONS OF THE GRAVITY FIELD AND EARTH PRECESSION-NUTATION} 
\vspace*{1cm} 

\noi \hspace*{1.5cm} G. BOURDA, N. CAPITAINE  \\ 
\noi \hspace*{1.5cm} SYRTE - UMR8630/CNRS, Observatoire de Paris  \\ 
\noi \hspace*{1.5cm} 61 avenue de l'Observatoire - 75014 Paris, FRANCE \\ 
\noi \hspace*{1.5cm} e-mail: Geraldine.Bourda@obspm.fr \\ 


\vspace*{0.7cm} 

\noi {\large 1. INTRODUCTION} 

\ssk 

Due to the accuracy now reached by space geodetic techniques, the temporal variations of a few Earth gravity field coefficients can be determined. Such variations result from Earth oceanic and solid tides, as well as from geophysical reservoirs masses displacements and post-glacial rebound. They are related to variations in the Earth's orientation parameters through their effect in the inertia tensor. 
We use (i) time series of the spherical harmonic coefficients $C_{20}$ ($C_{20}=-J_2$) of the geopotential and also (ii) $\Delta C_{20}$ models for removing a part of the geophysical effects. The series were obtained by the GRGS (Groupe de recherche en G\'{e}od\'{e}sie Spatiale, Toulouse) from the orbitography of several satellites (e.g. LAGEOS, Starlette, CHAMP) from 1985 to 2002 (Biancale et al., 2000).
In this preliminary approach, we investigate how these geodesic data can influence precession-nutation results.

\vspace*{0.7cm} 


\noi {\large 2. DATA AND METHOD} 

\ssk 
 
From the $C_{20}$ variation series, we can derive the corresponding variations of the dynamical flattening $H$, according to : $ \Delta H = -M ~R_e^2 ~\frac{\Delta C_{20}}{C} $, where $M$ is the mass of the Earth, $R_e$ its mean equatorial radius and $C$ its principal moment of inertia.
The $\Delta H$ series obtained in this way are mostly composed of an annual, semi-annual and 18.6-year terms. In order to investigate the influence of the variations in dynamical flattening on the precession-nutation, we integrate the following precession equations (Williams 1994, Capitaine et al. 2003 = P03) based on the observed $\Delta H$ series~: 
\begin{eqnarray}\label{eq:prec}
\sin \omega_A \frac{d\psi_A}{dt}   & = & \left( r_{\psi} ~\sin \epsilon_A \right) ~\cos \chi_A - r_{\epsilon} ~\sin \chi_A    \\
              \frac{d\omega_A}{dt} & = & r_{\epsilon} ~ \cos \chi_A + \left( r_{\psi} ~\sin \epsilon_A \right) ~\sin \chi_A  \nonumber
\end{eqnarray}
where $r_{\psi}$ and $r_{\epsilon}$ are the total contributions to the precession rate, respectively in longitude and obliquity, depending on the factor $H$.

\vspace*{0.7cm} 


\noi {\large 3. COMPUTATION AND RESULTS} 

\ssk  
 
We use the precession equations (\ref{eq:prec}) and the software GREGOIRE (Chapront, 2003), together with the $\Delta C_{20}$ data, to compute the effects in precession nutation. We find differences in the coefficients of the polynomial development of the precession angle $\psi_A$, depending on the $\Delta C_{20}$ contribution and the $J_2$ rate implemented ($J_2$ rate = $\dot{J}_2$). The results are composed of a polynomial part and a periodic part (i.e. Fourier and Poisson terms) discussed in the next paragraph. 
The effect due to the $J_2$ rate (i.e. effect on the $t^2$ term of $\psi_A$) can be taken into account using a series from 1985 to 1998 (Bourda and Capitaine, 2004). In Table \ref{tab:results}, our results rely on $\Delta C_{20}$ data from 1985 to 2002 and then do not take into account this effect.

\begin{table}[h]
\begin{minipage}{1.\linewidth}
\caption{{\small Polynomial expression for $\psi_A$ (up to degree 3) : (1) P03 and (2) Difference of our computation (influence of the $\Delta C_{20}$ \textit{residuals}, obtained with various $H$ constant parts) with respect to P03.}}\label{tab:results}
\begin{center}
\begin{tabular}{l|l|rrr}
\multicolumn{2}{c|}{}                                                 & {\small $t$}                   & {\small $t^2$}               & {\small $t^3$} \\
\hline
\multicolumn{2}{l|}{{\small (1) \textit{P03}}}                        & {\small \textit{5038".481507}} & {\small \textit{-1".079007}} & {\small \textit{-0".001140}} \\
\hline
{\small  (2) Difference of our}& {\small geodetic $H$ constant part}  & {\small 0".413188}             & {\small-0".001667}           & {\small $0".2~10^{-6}$}      \\
{\small ~~~~~computation w.r.t P03} & {\small VLBI $H$ constant part} & {\small 0"}	               & {\small-0".001579}           & {\small  $0".5~10^{-6}$}     \\
\hline
\end{tabular}
\end{center}
\end{minipage}

\hfill
\ssk

\begin{minipage}{1.\linewidth}
\caption{{\small Periodic contribution for the $t^0$ term of $\psi_A$, for different $\Delta C_{20}$ contributions; in microarcseconds.}} \label{tab:results_bis}
\begin{center}
\begin{tabular}{l|l|rr}
                                    & {\small Period}      &    {\small cos}       &   {\small sin}  \\
\hline
{\small Residuals}                  & {\small Annual}      &     {\small -1}       &    {\small  1}  \\
        			    & {\small Semi-annual} &     {\small  -}       &     {\small 1}  \\
				    \cline{2-4}
{\small Solid Earth tides}	    & {\small 18.6-yr}     &     {\small -2}       &   {\small 120}  \\
\hline
{\small TOTAL}			    & {\small 18.6-yr}     &     {\small  4}       &   {\small 105}  \\
\hline
\end{tabular}
\end{center}
\end{minipage}
\end{table}


\noi {\large 4. DISCUSSION} 

\ssk 

The precession rate (i.e. term in $t$ in the $\psi_A$ development) derived from the $C_{20}$ obtained by space geodetic techniques is smaller than the one obtained by VLBI (see Table \ref{tab:results}). The difference is about 400 mas/c, i.e. $\simeq 10^{-4} ~\times$ the precession rate value (this corresponds to a constant part of $-2.6835 ~10^{-7}$ in the $H$ value). Dehant and Capitaine (1997) already mentioned such a discrepancy relative to the IAU 1976 precession. Considering an error of about $10^{-10}$ in the $\Delta C_{20}$ data, we deduce an error of about 0.5 mas/c in the precession constant, which means that the difference obtained above is significant. In the future, several causes for this discrepancy will be investigated, such as the effect of the violation of hydrostatic equilibrium.

Then, the $H$ variations coming from the \textit{residuals} (i.e. $\Delta C_{20}$ data without atmospheric, oceanic tides or solid Earth tides $\Delta C_{20}$ models) observed by space geodetic techniques involved effects on the precession angle of about 1 $\mu$as or less (see Table \ref{tab:results_bis}). We also observed that the oceanic and atmospheric contributions were negligible. The principal periodic change, is due to the $\Delta C_{20}$ solid Earth Tides 18.6-yr variation, and is about 120 $\mu$as (in sine). 

For further studies, the Earth model has to be improved by considering (i) a refine Earth model, with core-mantle couplings and (ii) a reliable $J_2$ rate value.

\vspace*{0.7cm} 


\noi {\large 5. REFERENCES}  

{\leftskip=5mm 

\parindent=-5mm 

\ssk 

Biancale, R., Lemoine, J.-M., Loyer, S., Marty, J.-C., Perosanz, F., 2002, private communication of the C20 data.

Bourda, G., Capitaine, N., 2004, "Precession nutation and space geodetic determination of the Earth's variable gravity field", submitted to \aa.

Capitaine, N., Wallace, P.T., Chapront, J., 2003, "Expressions for the IAU 2000 precession quantities", \aa ~{\bf 412}, pp. 567-586.

Chapront, J., 2003, "Gregoire software", Notice, Paris Observatory.

Dehant, V., Capitaine, N., 1997, "On the precession constant : values and constraints on the dynamical ellipticity; link with Oppolzer terms and tilt-over-mode", \cmda ~{\bf 65}, pp. 439-458.

Williams, J.G., 1994, "Contributions to the Earth's Obliquity rate, Precession and Nutation",~\aj~{\bf 108 (2)},~pp.~711-724.

} 


\end{document}